\documentclass[twocolumn,prd,nofootinbib,aps,prl,floats,floatfix,superscriptaddress,amsmath,amssymb,longbibliography,secnumarabic]{revtex4-1} %
\usepackage[final]{graphicx}
\usepackage{hyperref}
\usepackage{amsmath}
\usepackage{bbm}
\usepackage{amsfonts}
\usepackage{amssymb}
\usepackage{latexsym}
\usepackage{graphicx}
\usepackage[english]{babel}
\usepackage{multirow}
\usepackage{float}
\usepackage{url}
\usepackage{slashed}
\usepackage{xcolor} 
\usepackage[utf8]{inputenc}
\usepackage{verbatim}
\usepackage{stackengine}

\def\sfrac#1#2{{\textstyle{#1\over #2}}}
\newcommand{\be}{\begin{equation}}
\newcommand{\ee}{\end{equation}}
\newcommand{\ba}{\begin{array}}
\newcommand{\ea}{\end{array}}
\newcommand{\bea}{\begin{eqnarray}}
\newcommand{\eea}{\end{eqnarray}}

\newcommand{\nn}{\nonumber}

\usepackage[utf8]{inputenc}

\begin{document}
\title{Nonabelian Kinetic Mixing in a Confining Phase}

\author{Gonzalo Alonso-{\'A}lvarez}
\email{gonzalo.alonso@utoronto.ca}
\thanks{ORCID: \href{https://orcid.org/0000-0002-5206-1177}{0000-0002-5206-1177}}
\affiliation{McGill University Department of Physics \& Trottier Space Institute, 3600 Rue University, Montr\'eal, QC, H3A 2T8, Canada}
\affiliation{Department of Physics, University of Toronto, Toronto, ON M5S 1A7, Canada}
\author{Ruike Cao}
\email{ruike\_cao@mail.ustc.edu.cn}
\thanks{ORCID: \href{https://orcid.org/0000-0002-4184-3560}{0000-0002-4184-3560}}
\affiliation{$^3$Department of Astronomy, School of Physical Sciences, University of Science and Technology of China, Hefei, Anhui 230006, China}
\author{James M.\ Cline}
\email{jcline@physics.mcgill.ca}
\thanks{ORCID: \href{https://orcid.org/0000-0001-7437-4193}{0000-0001-7437-4193}}
\affiliation{McGill University Department of Physics \& Trottier Space Institute, 3600 Rue University, Montr\'eal, QC, H3A 2T8, Canada}
\author{Karishma Moorthy}
\email{karishma.moorthy@mail.mcgill.ca}
\thanks{ORCID:\href{https://orcid.org/0009-0005-5995-3185}{0009-0005-5995-3185}}
\affiliation{McGill University Department of Physics \& Trottier Space Institute, 3600 Rue University, Montr\'eal, QC, H3A 2T8, Canada}
\author{Tianzhuo Xiao}
\email{tianzhuo.xiao@mail.mcgill.ca}
\thanks{ORCID: \href{https://orcid.org/0000-0002-1676-916X}{0000-0002-1676-916X}}
\affiliation{McGill University Department of Physics \& Trottier Space Institute, 3600 Rue University, Montr\'eal, QC, H3A 2T8, Canada}

\begin{abstract}
\noindent Dark matter from a hidden sector with SU($N$) gauge symmetry can have a nonabelian kinetic mixing portal with the standard model. 
The dark photon becomes massive in the confining phase without the need for spontaneous symmetry breaking.
Depending on the particle content of the dark sector, there can be two or more composite vectors that get kinetic mixing through a heavy mediator particle $X$.  
This provides a model of composite dark photons giving a portal for direct detection of dark baryons.
Avoiding exotic charged relics requires additional couplings allowing $X$ to decay to dark quarks and standard model fields, leading to further portals between the dark matter and the standard model. 
We comprehensively study the constraints on such models from colliders, rare decays, direct detection, and big bang nucleosynthesis.
\end{abstract} 
\maketitle

\section{Introduction}

\begin{figure}[t]
\begin{center}
 \includegraphics[scale=0.3]{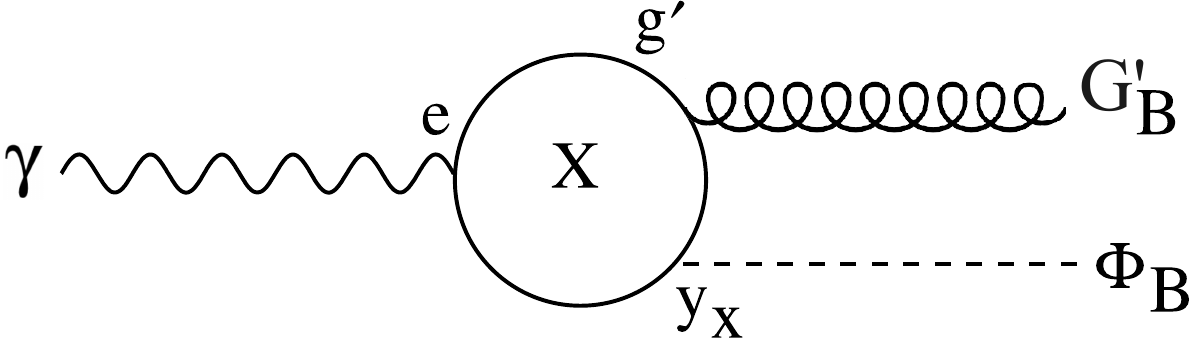}
 \caption{Nonabelian kinetic mixing from integrating out a heavy $X$ particle that carries electric charge and hidden SU($N)'$
 charge.}
 \label{fig:kmix}
\end{center} 
\end{figure}

Hidden sectors have become a rich arena for dark matter model
building \cite{Pospelov:2007mp,Arkani-Hamed:2008hhe}.  By virtue of portal interactions with the standard model (SM), they are hidden rather than being invisible.  For hidden sectors including a U$(1)'$ gauge interaction, kinetic mixing $\epsilon F_{\mu\nu}'
F^{\mu\nu}$ between the dark photon and the SM photon is a possible
portal, which is induced at one loop by integrating out a heavy particle $X$ that carries both kinds of charges \cite{Holdom:1985ag}.  If the dark gauge group is nonabelian,
such as SU($N)'$, such mixing is forbidden by gauge invariance, but higher dimensional versions such as
\be
    {\Phi_A\over M}\, {G'_A}^{\!\!\mu\nu} F_{\mu\nu}
    \label{nakm}
\ee
are possible~\cite{Arkani-Hamed:2008hhe,Chen:2009dm,Chen:2009ab} if there is an adjoint scalar field
$\Phi$ that also couples to 
$X$ in the loop,\footnote{One could alternatively have a fundamental representation scalar $\Psi$ leading to the dimension-6 operator
$M^{-2}\,(\Psi^\dagger T^A \Psi)\, G_A^{\mu\nu} F_{\mu\nu}$.} see Fig.~\ref{fig:kmix}.  If the gauge symmetry is broken by $\Phi$ getting a vacuum expectation value, then some linear combination of the dark gauge fields become a massive
photon, and Eq.\ (\ref{nakm}) reduces to the abelian case
with $\epsilon\sim \langle\Phi\rangle/M$.  This is the assumption that has been made in previous works that studied the phenomenology of nonabelian kinetic mixing \cite{Chiang:2013kqa,Cline:2014kaa,Cheung:2014tha,Barello:2015bhq,Choquette:2015mca,Carone:2018eka,Elahi:2020urr,Ko:2020qlt,Nomura:2021aep,Rizzo:2022jti,Zhou:2022pom,Cheng:2022aau}.

In the present work, we instead consider the situation when the SU($N)'$ symmetry remains unbroken and confines at a scale $\Lambda$.  Then the dark gluon $G'_\mu$
forms a bound state with the scalar $\Phi$, giving rise to a vector
$\tilde A^\mu$ whose mass originates from confinement rather than
symmetry breaking. 
This provides a model of a composite dark photon with kinetic mixing~\cite{Chiu:2022bni}.
In this work, we proceed to derive constraints on such models
from direct detection and collider searches, assuming that the dark matter is a baryon-like state of the SU($N)'$ sector.  

In fact, one quickly realizes that similar phenomenology can arise even in the limit where $m_\Phi\to\infty$ so that the $\tilde A^\mu$ decouples, since there are two additional vector states that acquire kinetic mixing with the photon via the heavy mediator $X$. These are the vector
meson $\omega^\mu$ that is a $Q\bar Q$ bound state,
and the $1^{--}$ vector glueball ${\cal G}^\mu$.
Here we also study the phenomenology of these states and delineate the regimes in which one of the three vectors dominates in the direct detection signal.

The cosmological history of the paradigm at hand is particularly rich. 
Notably, the $X$ particle would be stable in the simplest models, which is strongly constrained by searches for charged relics. 
To avoid this, we study two renormalizable extensions of the model that allow $X$ to decay into dark matter plus standard model particles. 
These additional interactions necessarily introduce additional portals for direct detection and collider searches that we thoroughly consider. 
Since the dark mesons can become fairly long lived, their late-time decay into SM states can impact the formation of light elements in the early universe. 
This leads to constraints from Big Bang Nucleosynthesis (BBN) that we derive in this work.

The dark baryon is stable due to its conserved number, and it can be an asymmetric dark matter candidate.  We will not be concerned with the mechanism of producing the asymmetry for it to have the right relic density, but rather assume that such a mechanism exists.  Because of its strong interactions, it is generic that the symmetric component will annihilate to negligible levels, as must happen for it to be asymmetric dark matter \cite{Graesser:2011wi}.
For example, if the annihilation cross section into dark mesons is geometric, $\sigma \sim 4\pi/\Lambda^2$, the symmetric component is 
exponentially suppressed as long as $\Lambda \lesssim 10^5\,$GeV, which is well within the range considered in this paper.

This paper is organized as follows.
We start by defining the minimal models that incorporate nonabelian kinetic mixing with composite dark matter candidate in Section~\ref{sec:models}. 
These are mapped onto a low energy effective description valid below the confinement scale in Section~\ref{sec:left}, to make contact with direct detection experiments
and cosmological constraints. 
In Section~\ref{sec:coll}, we derive constraints on the models from 
collider searches for the heavy $X$ mediator.
Constraints arising from direct searches for dark matter are presented in Section \ref{ddsect}, and lepton flavor violation
searches are discussed in Section \ref{sec:lfuv}. 
Limits on possible long-lived states from BBN are studied in
Section \ref{sec:cosmo}. 
We summarize and conclude in Section \ref{sec:conc}.

\section{Models}
\label{sec:models}
For simplicity, we consider there to be a single flavor of
dark quarks $Q^a$ transforming in the fundamental representation
of SU($N)'$, having bare mass $m_Q$. The $Q$ number is protected by a global dark baryon symmetry, and it is assumed that the $(Q)^N$ baryonic bound state is the stable dark matter candidate.

There may be an adjoint scalar
$\Phi^A$, inspired by the initial model of nonabelian kinetic mixing~\cite{Chiu:2022bni}, that interacts with $Q^a$ and $X^a$ via
\be
    y_q\, \bar Q \Phi Q +\left\{\begin{array}{cc} y_x \bar X \Phi X, & \hbox{fermionic $X$}\\
        \mu_x X^\dagger \Phi X,& \hbox{scalar $X$}
        \end{array}\right.,
        \label{yuk}
\ee
where $\Phi = T^A\Phi^A$ is the matrix form of the field in color
space.
The SU($N)'$ confinement scale is denoted by $\Lambda$, and its running coupling by $g'$.
The heavy mediator $X^a$ with mass $m_X$ is assumed to be in the fundamental representation.  As indicated in Eq.\ (\ref{yuk}), there is a choice as to whether $X$ is a Dirac fermion or a complex scalar, leading to the two classes of models that we consider.

If $\Phi^A$ gets a VEV, the diagram of 
Fig.\ \ref{fig:kmix} leads to kinetic mixing
of a linear combination of $G^A$ with the photon.  
In contrast to U($1)'$ hidden sectors, the loop integral is finite, and it gives rise to the coefficient
$1/M$ of Eq.\ (\ref{nakm})
\bea
{1\over M} ={e g'\over 16\pi^2}\times \left\{\begin{array}{cc} y_{x}/{m_{X}}, & \hbox{Dirac $X$}\\
\mu_{x}/2{m_{X}^2}, & \hbox{scalar $X$}
                                    \end{array}
                                    \right..
\eea  
On the other hand, if the SU($N$)$'$ remains unbroken, the kinetic mixing applies to a vector bound state $\tilde A$ of $\Phi^A$ and $G^A$, and the dimensionless kinetic mixing parameter arises from the product of $1/M$ and the decay constant $f_{\tilde A} = \langle 0|\Phi^A G^A|\tilde A'\rangle \sim \Lambda$.

To avoid charged relics, $X$ must decay into $Q$ and standard model particles.  At the renormalizable level, the possible couplings are limited, depending upon the spin of $X$. The operators allowed by gauge invariance are 
\be
    \lambda\, \bar Q H X \quad {\rm or}\quad \lambda_i\, \bar Q X e_{R,i}
    \quad +\quad {\rm H.c.}
\label{xdecay}
\ee
respectively, for Dirac or scalar $X$.  In the first case $X$ must be a vectorlike SU(2)$_L$ doublet to combine with the SM Higgs doublet $H$.  In the second, the coupling is to $i$th generation right-handed charged leptons, $e_{R,i}$.  This determines the weak hypercharge and electric charge(s) of $X$
in each case.  For Dirac $X$, the doublet members are charged as
$X = (X^0,\, X^+)^T$, allowing for the decays $X^+\to W^+ X^0$
followed by $X^0\to h Q$.  For the scalar, $X$ has charge $+1$ and decays to $Q$ plus charged leptons.

In summary, the ultraviolet ingredients of the hidden sector are relatively
simple: the nonabelian gauge fields, three kinds of matter fields, $Q$, $\Phi$ and $X$, and three kinds of Yukawa couplings, $y_x$ or $\mu_x$, $y_q$, and $\lambda$ or $\lambda_i$.  The weak hypercharge of $X$ is determined by its spin: $Y=+1/2$ for fermionic $X$ or $Y=1$
for bosonic $X$.  We use the convention $Y=Q-T_3$.

\begin{table}[t!]
\begin{tabular} {| c | c | c |c|}
\hline
Particle & Constituents & $J^{P(C)}$ & Mass\\
\hline
    $B$ & $Q^N$ & $(N/2)^+$  & $N(m_Q + \Lambda)$\\
\hline
    ${\tilde A}$  & $G'\Phi$ & $1^{--}$ &  $m_\Phi + \Lambda$ \\
     ${{\cal G}^\mu}$ & $G'G'$ & $1^{--}$ & $9\Lambda$ \\
    ${{\cal G}_0^\mu}$ & $G'G'$ & $1^{+-}$ & $7\Lambda$ \\
    $\omega^\mu$ & $Q\bar Q$ & $1^{--}$ &  $2(m_Q + \Lambda)$ \\
\hline
    $\eta$ & $Q\bar Q$ & $0^{-+}$ &  $2 m_Q + \Lambda$ \\
    ${\cal G}_0$ & $G'G'$ & $0^{++}$ &  $4\Lambda$ \\
    ${\cal G}$ & $G'G'$  & $0^{-+}$ &  $6\Lambda$ \\
    ${\eta_\Phi}$ & $Q\Phi\bar Q$ & $0^{-+}$ & $2 m_Q + m_\Phi + \Lambda$ \\
   ${S}$ & $\Phi\Phi$ & $0^{++}$ &  $2m_\Phi + \Lambda$\\
\hline
\end{tabular}
\caption{Bound states in the low-energy effective theory, classified by spin, parity, and charge conjugation, along with quark model or lattice estimates for their masses.}
\label{tab1}
\end{table}

\section{Low energy effective theory}
\label{sec:left}
Having defined the model in terms of the fundamental constituents,
we need to describe it below the confinement scale, where direct detection and BBN constraints will be applied.  The low-energy theory contains a dark baryonic state $B \sim (Q)^N$,
the composite vector $\tilde A^\mu \sim G_B^\mu\Phi_B$, a pseudoscalar meson $\eta \sim Q\bar Q$, and a vector meson $\omega^\mu\sim \bar Q\gamma^\mu Q$, where we have indicated the interpolating fields of the fundamental theory.   In addition, there are glueballs ${\cal G} \sim GG$ and scalar balls $\eta_\Phi \sim \bar Q\Phi Q$,
$S\sim \Phi\Phi$.  Schematically, the bound state masses are estimated in Table \ref{tab1},
using quark model estimates or lattice
QCD \cite{Chen_2006,Lucini_2010}.  There is no chirally suppressed meson mass since the global U(1)$_A$ quark flavor symmetry is anomalous. 

\subsection{Radiative decay operators}

In our model, the dark baryon $B$ is the dark matter candidate, while the various mesonic bound states are rendered unstable by the portal interactions introduced above.
Their main phenomenological interest is that they must decay fast enough to satisfy constraints from Big Bang Nucleosynthesis.  We will consider those constraints in detail in section \ref{sec:cosmo}.  Here, we estimate the effective interactions enabling the decays.    
Neglecting mass mixing, the scalar and pseudoscalar states can decay into two photons through the effective interactions
\be
   {\cal L} \ni \left( {\eta\over \Lambda_\eta} + 
   {{\cal G}\over \Lambda_{\cal G}}\right) F\tilde F 
    +\left({ {\cal G}_0\over \Lambda_{{\cal G}_0} }
    +{S\over \Lambda_{S}}\right) F^2
    \label{meson-decays}
\ee
where $\Lambda_i$ are mass scales to be estimated below. 

\begin{figure}[t]
\centering
\includegraphics[width=0.3\linewidth]{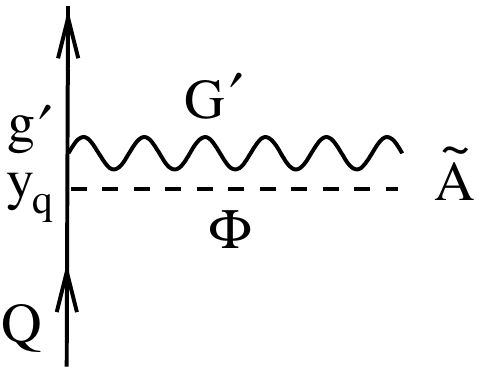}
\caption{Effective interaction of the dark quark $Q$ with the $\tilde A^\mu$ composite vector.
}
\label{fig:QAt}
\end{figure}

The diagram in Fig.\ \ref{fig:QAt} leads to mass mixing between the composite vectors $\tilde A^\mu$ and $\omega^\mu$.  The same diagram without the $G'$ line gives mixing between the  
pseudoscalar $\eta_\Phi$ and $\eta$ mesons.  Since the gauge coupling $g'\sim 1$ at the scales of the mesons,
the mixing mass squared is of order $y_q \Lambda^2$ for both systems.  Then
$\eta_\Phi$ can decay to two photons by its mixing with
$\eta$, and $\tilde A$ can decay like
$\omega\to \eta \gamma$ though its mixing with $\omega$.  The latter mixing angle is 
\be
    \theta_{\tilde A} \sim {y_q\Lambda^2\over
        m_{\omega}^2 - m_{\tilde A}^2}
        \label{thetaA}
\ee
with the masses given in Table \ref{tab1}.
In addition, $\tilde A$ can decay into SM fermions via its kinetic mixing. 

\begin{figure*}[t]
\begin{center}
\vskip-1cm
\centerline{\raisebox{-0.35cm}{\includegraphics[width=0.36\textwidth]{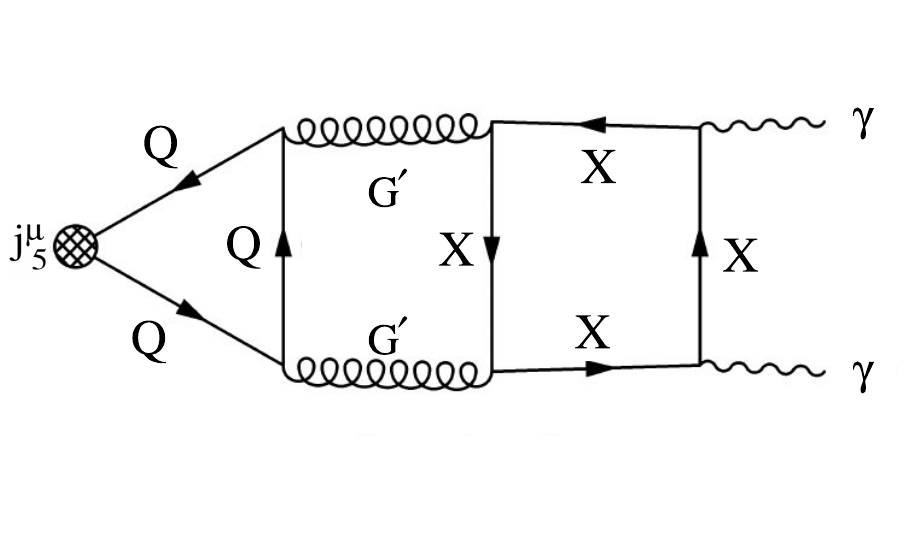}}
\includegraphics[width=0.31\textwidth]{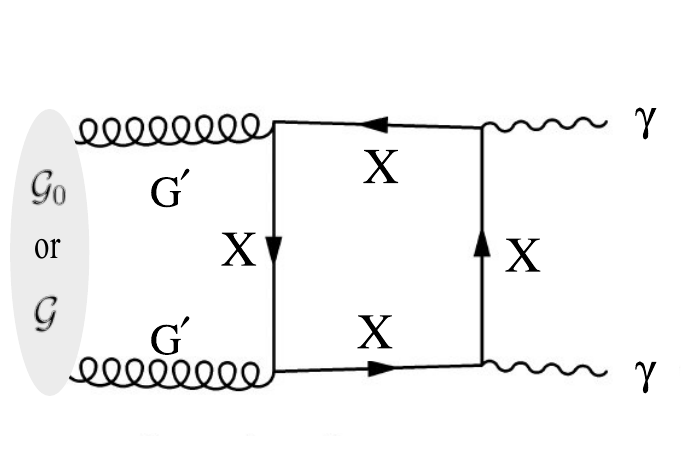}
\includegraphics[width=0.31\textwidth]{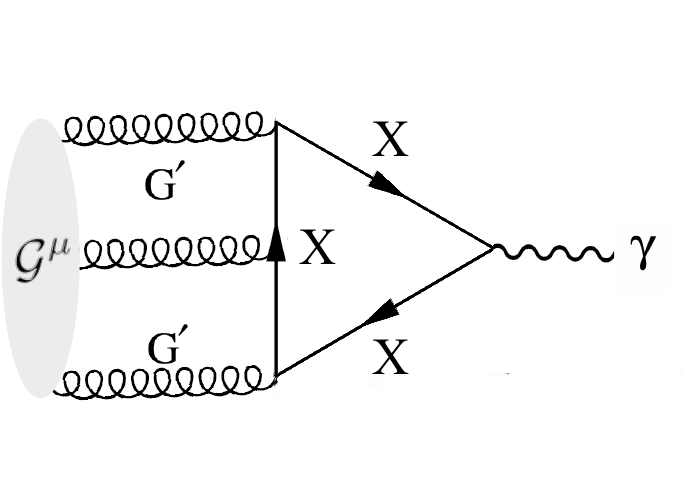}}
\vskip-0.5cm
 \caption{Left: diagram contributing to the decay of the dark meson $\eta$ into two visible photons.
 Center: diagram leading to glueball decays to two photons.  By replacing both gluon lines with
 $\Phi$, we get decays $S\to \gamma\gamma$.
 Right: diagram leading to kinetic mixing of the vector glueball with the photon.
 }
 \label{fig:eta_decay}
\end{center} 
\end{figure*}

The low-energy effective coupling for $\eta$ can be estimated from the first diagram of Fig.\ \ref{fig:eta_decay}. It makes a  contribution to the divergence of the dark axial vector current $j_5^\mu = \bar Q\gamma^\mu\gamma_5 Q$, that interpolates between $\eta$ and two photons. The value can be inferred from Ref.\ \cite{Anselm:1989gi}, which computed the same diagram in the fully Abelian case.  Correcting for the color factor, we estimate that
\be
    \partial_\mu j^\mu_5(q^2) \cong {N\alpha\alpha'^2\over 4\pi^3}
        \ln{m_X^2\over q^2} F\tilde F = 
        {f_\eta\over \Lambda_\eta} F\tilde F
        \label{divj5}
\ee
with $q^2 = m_\eta^2$ and $f_\eta \sim \Lambda$.
Here, $\alpha$ denotes the electromagnetic structure constant.
The gauge coupling $\alpha'$ should be evaluated at the scale $m_X$ here and below. For our purposes, the one-loop beta function \cite{Ryttov:2007cx} gives a sufficient estimate, $\alpha'(m_X)\sim 6\pi/[(11N-6)\ln(m_X/\Lambda)]$.

The second diagram of Fig.\ \ref{fig:eta_decay} was computed in Ref.\ \cite{Juknevich:2009ji}, giving rise to the effective operators
\be
    {\alpha \alpha'\over m_X^4} \left(
        \sfrac{1}{45}{\rm tr}[{G'\tilde G'}] F\tilde F
    + \sfrac{1}{60}{\rm tr}[{G'G'}] F^2\right)\,.
    \label{glueball_dim_8}
\ee
in the case of fermionic $X$.  For bosonic $X$,
in analogy to scalar QED, we estimate an extra factor of $-1/2$ for each diagram \cite{Huet:2011kd}.
To relate it to the hadronic effective description
(\ref{meson-decays}), we must estimate the decay constants
such as $\langle 0|{\rm tr}[{G'\tilde G'}]|{\cal G}\rangle = f_{\cal G} \sim \Lambda^3$. 
Then
\bea
\Lambda_{\cal G} &\sim& 45\, m_X^4/( \alpha'\alpha\,\Lambda^3)\,,\nn\\ 
\Lambda_{{\cal G}_0} &\sim& 60\, m_X^4/(\alpha'\alpha \,\Lambda^3)\,.
\label{glueball mass}
\eea
We further estimate that $\Lambda_{S}\sim
\Lambda_{{\cal G}_0}$.  

The radiative decay rates for these states  have been computed (up to the decay constants, which we estimate dimensionally as $\Lambda^3$) in Ref.\ \cite{Juknevich:2009ji}.  The vector glueball decays are also treated there; they include
${\cal G}_0^\mu\to {\cal G}_0\gamma$ and 
${\cal G}^\mu\to {\cal G}\gamma$.
For the lightest $0^{++}$ state, 
\be
    \Gamma_{{\cal G}_0\to\gamma\gamma} \sim
        {2(N^2-1)\Lambda^3\over
        \pi\, \Lambda_{{\cal G}_0}^2}\,.
\label{gbdecay}
\ee
The $1^{+-}$ and $1^{--}$ vector glueballs can radiatively decay to the ground state with rate
\be
    \Gamma_{{\cal G}_{(0)}^\mu\to \gamma {\cal G}_0}
    \sim (100-400)\,{\alpha\alpha'^3\,\Lambda^9\over 24\pi\, m_X^8}\,,
\ee
where the larger coefficient applies to the heavier parent particle.

\begin{table}[t!]
\begin{tabular} {| c | c |c | c |}
\hline
particle & $\epsilon_v$ (Dirac $X$) & $\epsilon_v$ (scalar $X$) & $g_v$ \\
\hline
$\tilde A^\mu$ & ${\sqrt{\alpha\alpha'}y_x\Lambda^{\phantom{|^|}}\over 4\pi m_X}$
& ${\sqrt{\alpha\alpha'}\mu_x\Lambda\over 8\pi {m_X^2}_{\phantom{|}}}$ 
& ${y_q\Lambda\over \sqrt{N}( m_Q + \Lambda)}$\\
\hline
$\omega^\mu$ & ${e g_2^2 \lambda^2 v^2 \Lambda^{\phantom{|^|}}\over 16\pi^2 m_X^3}$ & 
${e\sum_i \lambda_i^2\,m_{\ell_i} \Lambda^{\phantom{|^|}}\over 16\pi^2{m_X^2}_{\phantom{|}}}$
& $4\pi$ \\
\hline
${\cal G}^\mu$ & $ \sfrac23\sqrt{\alpha\alpha'^3}\left(\Lambda\over m_X\right)^4$  &
$ \sfrac13\sqrt{\alpha\alpha'^3}\left(\Lambda\over m_X\right)^4$
& $4\pi\over \sqrt{N}$ \\
\hline
\end{tabular}
\caption{The low-mass composite vector states, their kinetic mixing parameters (depending upon the spin of the mediator $X$) and their vectorial coupling to the dark baryon current.}
\label{tab2}
\end{table}

\subsection{Kinetic mixing and couplings to DM}

The portal interactions (\ref{yuk}) further give rise to
glueball decays into Higgs bosons \cite{Juknevich:2009ji} or right-handed leptons, and kinetic mixing with ${\cal G}^\mu$. 
The latter comes from the right diagram in Fig.~\ref{fig:eta_decay} and gives rise to the effective operators \cite{Falkowski:2009yz}
\be
    {\alpha'^{3/2}\alpha^{1/2}\over m_X^4} F^{\mu\nu} \left(
    {14\over 45}{\rm tr}\, G'_{\mu\nu} G'_{\alpha\beta}G'^{\alpha\beta} -
    \frac19{\rm tr}\,{G'\!\!}_{\mu}^{\ \alpha}{G'\!\!}_{\alpha}^{\ \beta}G'_{\nu\beta}\right)\,.
\ee
These operators interpolate between the photon and the two vector glueballs (despite their opposite parity \cite{Jaffe:1985qp}); however parity forbids kinetic mixing with the $1^{+-}$ lowest vector glueball ${\cal G}^\mu_0$.   One can estimate that the ensuing effective kinetic mixing term is
\be
   \sfrac12 \epsilon_{\cal G}F_{\mu\nu}{\cal G}^{\mu\nu} \sim \frac13\alpha^{1/2}\alpha'^{3/2} \left(\Lambda\over m_X\right)^4 F_{\mu\nu}{\cal G}^{\mu\nu}
\ee
with the glueball field strength tensor. 
This exchange mediates scattering between the dark $B$ baryon and the SM proton, for which the effective coupling $g_{\cal G}$ between ${\cal G}^\mu$ and $B$ is required.  From large-$N$ counting rules \cite{Manohar:1998xv}, one finds that the  vector coupling of
${\cal G}^\mu$ is suppressed relative to that of  $\omega^\mu$, which does not scale with $N$. 
Using the $\omega$-nucleon coupling in QCD
\cite{Furnstahl:1996wv}, we estimate
\be
   \sqrt{N} g_{\cal G} \cong g_\omega \cong 4\pi\,.
   \label{gbcoupling}
\ee

For the $\tilde A^\mu$ composite vector, we estimate
the kinetic mixing and effective coupling to $\bar B\gamma^\mu B$ as
\be
    \epsilon_{\tilde A} \sim \alpha_x \alpha'^{1/2}
    \alpha^{1/2}\,{\Lambda\over m_X},\ \ 
    g_{\tilde A} \sim \frac{y_q \Lambda}{\sqrt{N}(m_Q+\Lambda)}\,.
\label{epsilon_eff}
\ee
where $\alpha_x = y_x/4\pi$ for Dirac $X$, and 
$\alpha_x = \mu_x/(8\pi m_X)$ for scalar $X$.
A diagram giving rise to $g_{\tilde A}$ is shown in Fig.\ \ref{fig:QAt}, where $g'\sim 1$ since it is evaluated at a scale $\sim \Lambda$.  Like the glueball coupling in Eq.\ (\ref{gbcoupling}),
it is suppressed at large $N$.

\begin{figure}[b]
\begin{center}
 \includegraphics[scale=0.3]{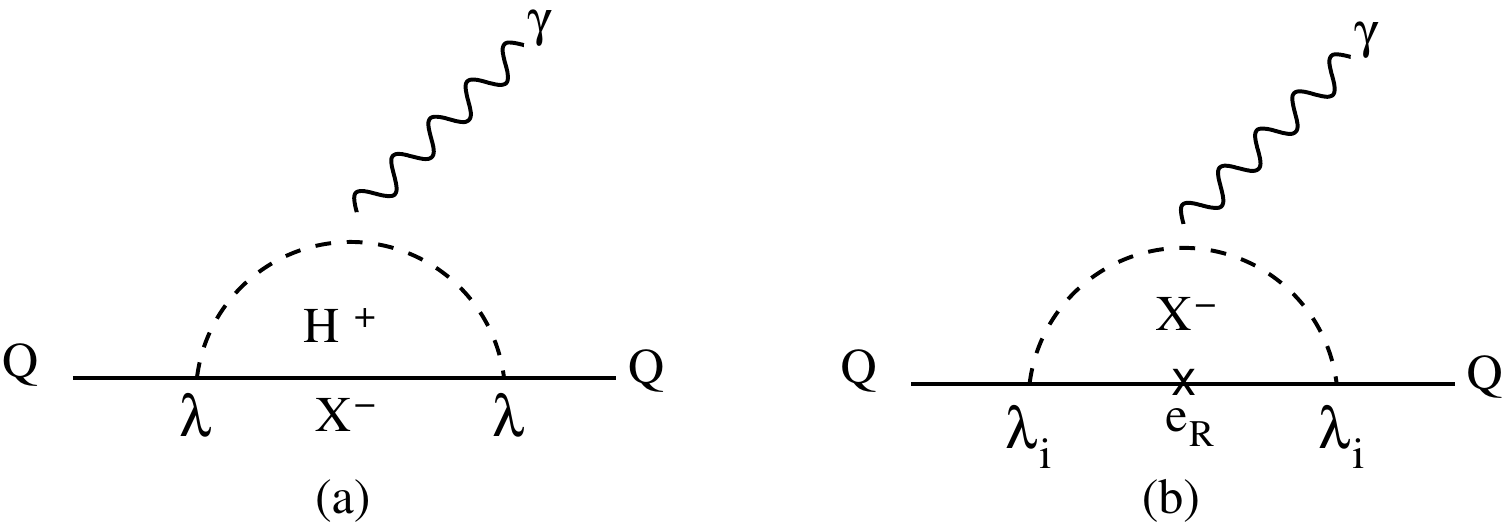}
 \caption{Diagrams generating dark baryon magnetic moment in the (a) fermionic $X$ and (b) scalar $X$
variants of the model.  In addition to diagram (a) which appears in `t Hooft gauge, there is a subdominant $W$ loop contribution.}
 \label{fig:mag}
\end{center} 
\end{figure}

The remaining light vector state is the meson
$\omega^\mu$, which can mix with 
$\tilde A^\mu$ through an off-diagonal mass term $\delta m^2 
 \omega^\mu\tilde A_\mu$ of order $\delta m^2 \sim  y_q \Lambda^2$.
It can acquire kinetic mixing with the SM hypercharge by virtue of the one-loop contribution to the $Q$
magnetic moment, from the diagrams in Fig.\ \ref{fig:mag}. The kinetic mixing between $\omega^\mu$ and the photon is interpolated by the
magnetic moment interaction, 
\be
    \langle \omega^\mu |\mu_Q \bar Q\sigma^{\rho\sigma}Q F_{\rho\sigma} |0\rangle \to 
     \epsilon_\omega \omega^{\rho\sigma} F_{\rho\sigma}\,,
\ee
where $\omega^{\rho\sigma}$ is the $\omega^\mu$ field strength.  One can then estimate that $\epsilon_\omega \sim \mu_Q\Lambda$, where $\mu_Q$
is the loop-generated quark magnetic dipole moment (MDM),
\be
\mu_Q \cong {e\over 16\pi^2 m_X^2}\left\{\begin{array}{cc}{\lambda^2 m_X+\theta^2 g_2^2 m_X}, & \hbox{Dirac $X$}\\
  \sfrac13{  \sum_i\lambda_i^2m_Q \ln{m_X\over m_{\ell_i}}}                                , & \hbox{scalar $X$}
\end{array}
\right..
\label{MDM}
\ee
For fermionic $X$, the second term arises from virtual $W$ exchange, while in the scalar $X$ result,
the log enhancement comes from the diagram where
$\gamma$ attaches to the lepton with mass $m_{\ell_i}$.

We summarize the predicted kinetic mixings and couplings to the dark baryon of the three vector states in table
\ref{tab2}.  These, along with the vector masses, are the relevant quantities for direct detection via nonabelian kinetic mixing, to be discussed in Section \ref{ddsect}.

\section{Collider constraints}
\label{sec:coll}
The SU(2)$_L$ charged mediator $X$ can be produced at high energy collider experiments.
Depending on its spin, different search strategies are best suited to look for it.

\begin{figure*}[t]
\centering
\centerline{\includegraphics[width=0.5\linewidth]{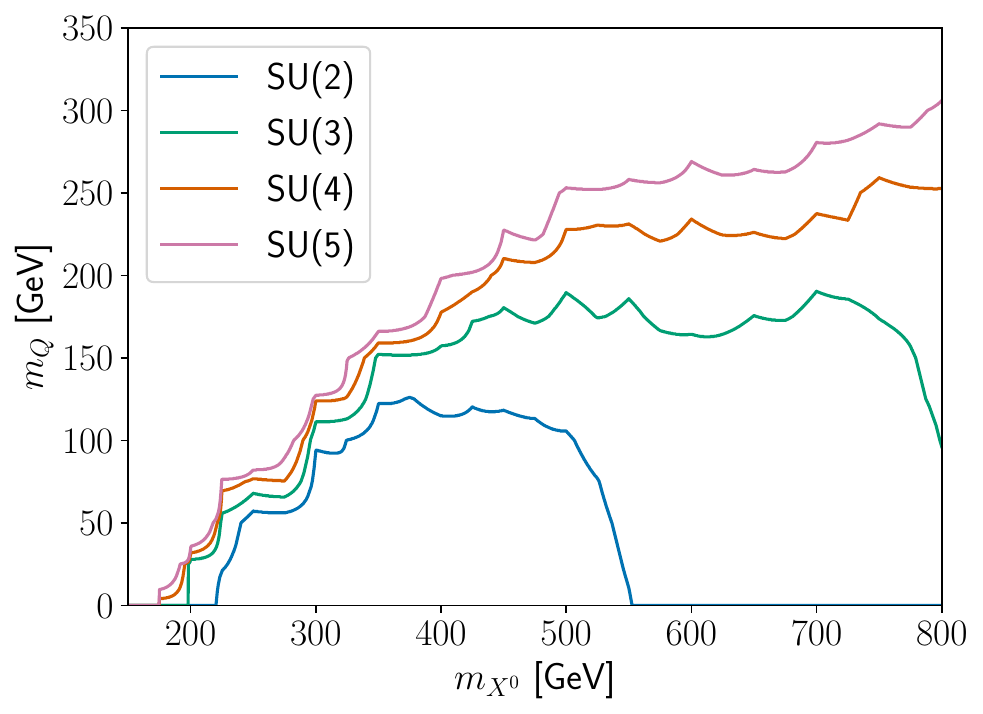}
\includegraphics[width=0.5\linewidth]{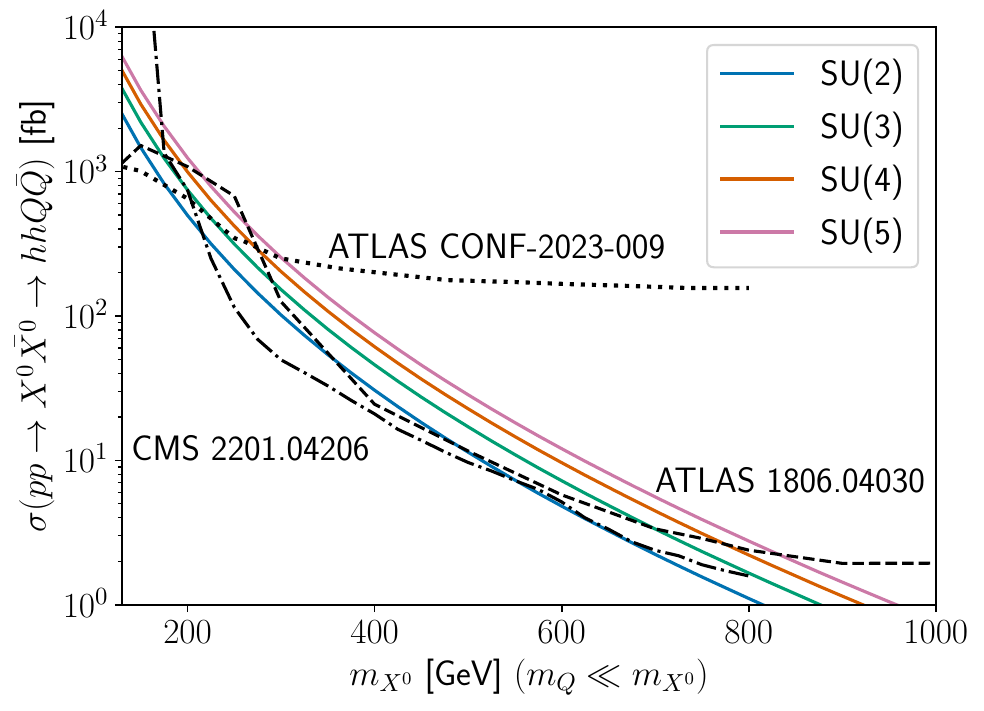}}
\caption{Constraints on the neutral component of the Dirac fermion $X$ mediator from LHC searches for Higgsino pair production decaying into a Higgs boson pair plus missing energy.
\emph{Left:} Recasted limits from~\cite{CMS:2022vpy} as a function of $m_{X^0}$ and the mass of its $Q$ decay product.
For different dark gauge groups, the regions below each line are excluded.
\emph{Right:} The black dashed and dash-dotted lines show the cross section upper limits from~\cite{ATLAS:2018tti,CMS:2022vpy,ATLAS:2023rie} as a function of $m_{X^0}$ in the limit where $Q$ is assumed to be much lighter.
The colored lines represent the theoretical predictions for different confining gauge groups. 
}
\label{fig:LHC_X0}
\end{figure*}

\subsection{Fermion mediator}
The case where $X$ is a Dirac fermion with $Y=+1/2$ leads to phenomenology that is similar to that of a supersymmetric Higgsino.
Via the operator $\lambda \bar{Q}HX$, the neutral component will decay $100\%$ of the time into $X^0 \rightarrow h + Q$, as long as the decay is kinematically allowed.
We assume that $\lambda$ is sufficiently large so that the decay is prompt.

With these assumptions, $X^0$ is constrained by searches looking for pair-produced neutralinos decaying into a Higgs boson and a neutral LSP.
In most of the parameter space, the strongest bounds are placed by the recent CMS search~\cite{CMS:2022vpy}.
At masses below $\sim 200$~GeV and above $800$~GeV, the ATLAS searches~\cite{ATLAS:2018tti,ATLAS:2023rie} improve the limits, assuming that the stable neutral particle that carries away the missing energy is massless.

To recast the aforementioned analyses for our model, we implement the relevant particles and interactions in \texttt{FeynRules}~\cite{Alloul:2013bka} and calculate the leading-order (LO) $pp\rightarrow X^0 \bar{X}^0$ production cross-section using \texttt{MadGraph5\_aMC@NLO}~\cite{Alwall:2014hca}.
The next-to-leading-order (NLO) and next-to-leading-log (NLL) effects~\cite{Fuks:2012qx,Fuks:2013vua} are incorporated by correcting the LO cross-section by a K-factor of $1.4$.
The total cross-section is boosted by the dark color multiplicity of $X^0$, which depends on the dimension of the SU($N)'$ dark gauge group.

The resulting limits are shown in Fig.~\ref{fig:LHC_X0} and exclude mediators with masses below $800$~GeV for the benchmark case of an SU(3) confining gauge group.
The left panel of Fig.~\ref{fig:LHC_X0} shows the exclusion arising from the CMS search~\cite{CMS:2022vpy}, as a function of $m_{X^0}$ and $m_Q$.
The kinks in the exclusion lines can be traced back to the binning used in the experimental analysis.
As expected, the limits are strongest for $m_Q = 0$ and vanish or significantly degrade for $m_Q\gtrsim m_h$.
Since we are not including hadronization processes in the dark sector, we use the parton-level mass $m_Q$ for the final state dark quarks.
This should be a good approximation when $m_Q\gtrsim\Lambda$.

The right panel shows the cross-section upper limits for the CMS~\cite{CMS:2022vpy} and ATLAS~\cite{ATLAS:2018tti,ATLAS:2023rie} searches for massless $Q$ (or $m_Q\ll m_h$).
Except for a small range of masses around $200$~GeV for a SU(2) gauge group, $m_{X^0}$ is constrained to be above $600$~GeV for SU(2), $800$~GeV for SU(3), and even higher masses for larger dark gauge groups.

To conserve dark SU(N) charge, the $Q\bar{Q}$ pair are produced with opposite dark color. Since the interaction is strong, color strings stretch between the two $Q$ particles.
This can lead to distinctive kinematic features in the missing energy spectrum of the events that could be used to further test the model in a dedicated search for hidden-valley type of dark sector models~\cite{Strassler:2006im,Han:2007ae}, but that we have neglected in the present analysis.

\begin{figure*}[t]
\centering
\centerline{\includegraphics[width=0.5\linewidth]{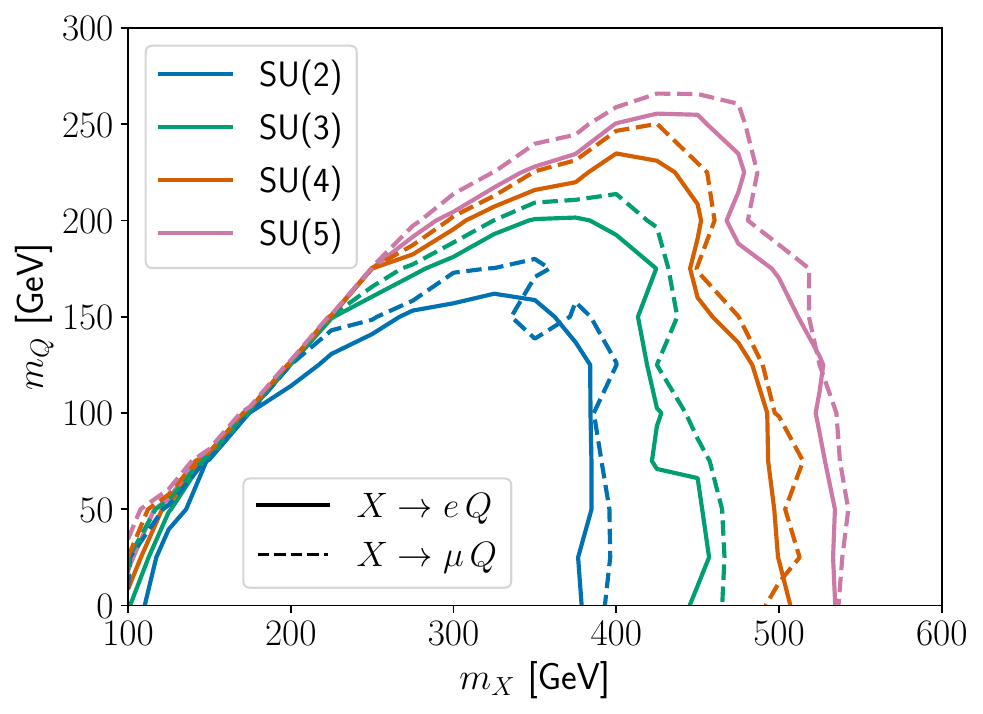}\includegraphics[width=0.5\linewidth]{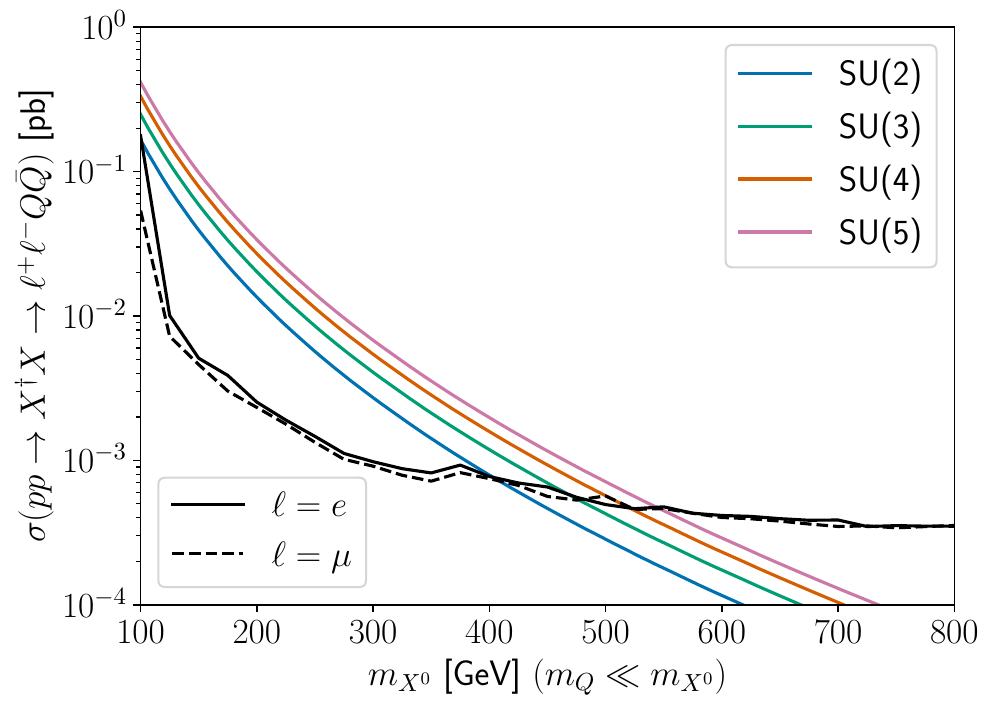}}
\caption{Constraints on the scalar $X$ mediator from LHC searches for slepton pair production decaying into an opposite-sign lepton pair plus missing energy.
\emph{Left:} Recasted limits from~\cite{ATLAS:2019lff} as a function of $m_{X}$ and the mass of its $Q$ decay product.
For different dark gauge groups, the regions below the correspondingly colored lines are excluded for dominant electron or muon decays, respectively.
\emph{Right:} The black solid and dashed lines display the cross section upper limits from~\cite{ATLAS:2019lff} as a function of $m_{X}$ in the limit of massless $Q$, for dielectron and dimuon final states, respectively.
The colored lines show the cross section predictions for different confining gauge groups.  
}
\label{fig:LHC_ScalarX}
\end{figure*}

The charged component of the SU(2) $X$ doublet decays via $X^+ \rightarrow W^+ X^0$.
The mass splitting due to radiative corrections between the charged and neutral components is~\cite{Thomas:1998wy}
\begin{equation}
    \delta m = m_{X^+} - m_{X^0} = 
    \frac{\alpha}{2}\, m_Z\, f\left(\frac{m_X^2}{m_Z^2}\right)
    - {(\lambda v)^2\over 2 m_X}\,,
    \label{dmxeq}
\end{equation}
where $m_X$ arises from the vector-like mass term $m_X\bar{X}X$ and $f$ is the loop function
\begin{equation}
    f(r) = \frac{\sqrt{r}}{\pi} \int_{0}^{1} \mathop{\mathrm{d}x} \, (2-x) \, \ln{\left[ 1 + \frac{x}{r(1-x)^2} \right]}\,.
\end{equation}
For $m_X \gg m_Z$, the loop-generated mass splitting asymptotes to $\delta m\simeq 355$~MeV.  This dominates over the
second contribution in (\ref{dmxeq}), from 
 $Q$-$X^0$ mixing, since direct detection limits (see Fig.\ \ref{fig:cs} below) constrain this contribution to be $\lambda v\theta/\sqrt{2}$ $\lesssim$ $O(10)$ MeV.
 For the masses of interest $m_X\gtrsim 100$~GeV, the leading decay channel for the charged component is $X^+\rightarrow \pi^+X^0$, with a rate
\begin{equation}
    \Gamma(X^+\rightarrow \pi^+X^0) = \frac{G_F^2}{\pi} \, |V_{ud}|^2 \, f_\pi^2 \, \delta m^3 \, \sqrt{1-\frac{m^2_{\pi^+}}{\delta m ^2}}\,,
\end{equation}
with $f_\pi \simeq 130$~MeV.
This corresponds to a lifetime
\begin{equation}
    c\tau \simeq 0.6\,\mathrm{cm}\,\left
( \frac{355\,\mathrm{GeV}}{\delta m} \right)^3 \, \left( 1 - \frac{m^2_{\pi^+}}{\delta m ^2} \right)^{-1/2},
\end{equation}
which is macroscopic but fairly short for collider searches.
This makes searches for these decays extremely challenging.
The current best limits are set by an ATLAS search for long-lived charginos~\cite{ATLAS:2022rme}, but the limits become weak for the small displacements of interest here, only excluding masses $m_X\lesssim 200$~GeV.
Since this is significantly smaller than the exclusion arising from searches of the neutral component of $X$, we do not attempt to perform a full recast of this search to our scenario.
Alternative search strategies have been proposed~\cite{Fukuda:2017jmk} that may help improve the bounds at the high-luminosity phase of the LHC.

\subsection{Scalar mediator}
In the scalar mediator case, $X$ is an SU(2)$_L$ singlet with charge +1 and decays to a right-handed lepton plus missing energy through the operator in Eq.~\eqref{xdecay}.
For pair produced mediators, this leads to a collider signature featuring missing energy and oppositely charged leptons.
This signal is reminiscent of that induced by sleptons in some supersymmetric extensions of the SM.
Searches for this final state have been performed at CMS~\cite{CMS:2020bfa,CMS:2022syk} and ATLAS~\cite{ATLAS:2019lff,ATLAS:2022hbt,ATLAS:2019lng,ATLAS-CONF-2023-029}.

We apply the ATLAS limits~\cite{ATLAS:2019lff} to our physics model using the publicly available \texttt{MADANALYSIS5}~\cite{Conte:2012fm,Conte:2014zja,Conte:2018vmg} recast of the analysis~\cite{DVN/EA4S4D_2020}.
Using \texttt{FeynRules}~\cite{Alloul:2013bka} and \texttt{MadGraph5\_aMC@NLO}~\cite{Alwall:2014hca}, we calculate the LO $pp\rightarrow X \bar{X}$ production cross-section.
In this case we do not apply any K-factor correction as the NLO+NLL effects are small for this process~\cite{Fiaschi:2018xdm}.
We also make use of the hadronization and detector simulation codes \texttt{PYTHIA8}~\cite{Sjostrand:2014zea} and \texttt{DELPHES3}~\cite{deFavereau:2013fsa}.

The result of the above process is shown in Fig.~\ref{fig:LHC_ScalarX}.
The left panel shows constraints in the $m_X$ vs.~$m_Q$ plane, assuming that $X$ decays 100\% of the time into either electrons or muons.
Limits on tau lepton decays are only slightly weaker~\cite{CMS:2022syk,ATLAS-CONF-2023-029}.
The right panel shows the constraints for the case where the dark quark mass is much smaller than $m_X$.
Depending on the dimension of the dark gauge group, we find a lower bound of $m_X\geq 400-600\,\mathrm{GeV}$.
As in the fermion mediator search, these searches do not exploit any distinctive kinematic features arising from the strong interactions among the final state dark quarks.

At LEP, X mediates the t-channel process $e^+e^- \rightarrow Q \bar{Q}$.
The L3 collaboration looked for single- and multi-photon events with missing energy~\cite{L3:2003yon}, which can occur when a photon is radiated off a charged particle in the previous process.
The observed rate for such events at LEP matches well the SM predictions due to $e^+e^-\rightarrow\nu\bar{\nu}\gamma$, and can thus be used to place bounds on our model.
Using \texttt{FeynRules}~\cite{Alloul:2013bka} and \texttt{MadGraph5\_aMC@NLO}~\cite{Alwall:2014hca}, we find the cross section for the new physics process to be
\begin{equation}
    \sigma(e^+e^- \rightarrow Q \bar{Q} \gamma) \simeq 8\times 10^{-4}\,\mathrm{pb}\, N \, \lambda_1^4 \left( \frac{1\,\mathrm{TeV}}{m_X} \right)^4
\end{equation}
at $\sqrt{s}=200$~GeV for photons with $p_T > 4$~GeV and $m_Q\ll m_X$.
Here, $N$ denotes the number of colors in the dark SU$(N)'$ gauge group.
In the total luminosity of $619\,\mathrm{pb}^{-1}$, 1898 single-photon events were observed compared with the SM expectation of $1905.1$ (see table 2 in~\cite{L3:2003yon}). Given that trigger and selection efficiencies are $\sim 70\%$, we place a rough limit by demanding the new physics events do not exceed 10, leading to
\begin{equation}
    \frac{m_X}{\lambda_1}\gtrsim 0.5\, N^{1/4} \,\mathrm{TeV}\,.
\label{LEP}
\end{equation}

\section{Direct detection}
\label{ddsect}
The framework of nonabelian kinetic mixing leads to dark matter direct detection signals through
exchange of composite vector bosons.
But in addition, there are several other portals that inevitably arise when the new interactions needed to avoid the generic heavy
charged relic problem are introduced.
We first consider the
direct detection signals arising from these extra portals, specific to the fermionic
or scalar $X$ models respectively, and then turn to the kinetically mixed
vector exchange. 
As we will se, the allowed values for the latter process depend upon couplings that are constrained by the former ones.

\begin{figure}[t]
\centering
\includegraphics[width=\linewidth]{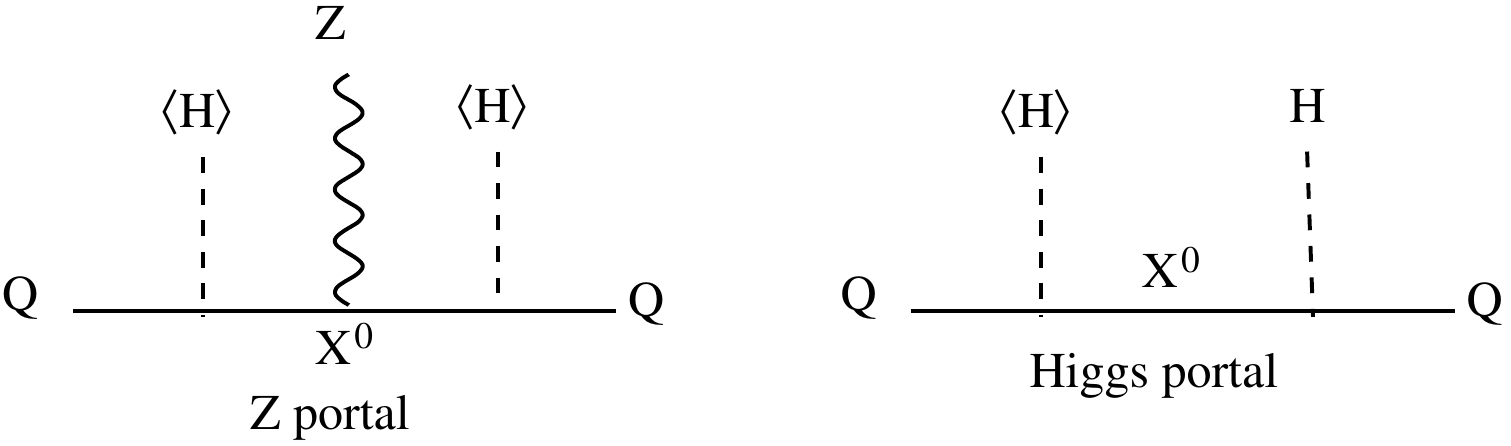}
\caption{Effective interactions of $Q$ with the $Z^0$ and Higgs boson mediating interactions of dark quark $Q$ with nucleons, for fermionic $X$ mediator.
}
\label{fig:portals}
\end{figure}

\subsection{Fermionic $X$ mediator}

In addition to the magnetic dipole operator 
generated by the diagram in Fig.\ \ref{fig:mag}, the fermionic $X$ mediator gives rise to tree-level Higgs and $Z$ exchange between dark baryons and nucleons, from the diagrams in Fig.~\ref{fig:portals}.  These arise from the mixing of $X^0$ and $Q$ at electroweak symmetry breaking,
from the operator (\ref{xdecay}).  The mixing angle is $\theta \cong \lambda v/(\sqrt{2}\, m_X)$ when $\theta \ll 1$ and $m_X\gg m_Q$.

\begin{figure}[t]
\begin{center}
 \includegraphics[width=0.5\textwidth]{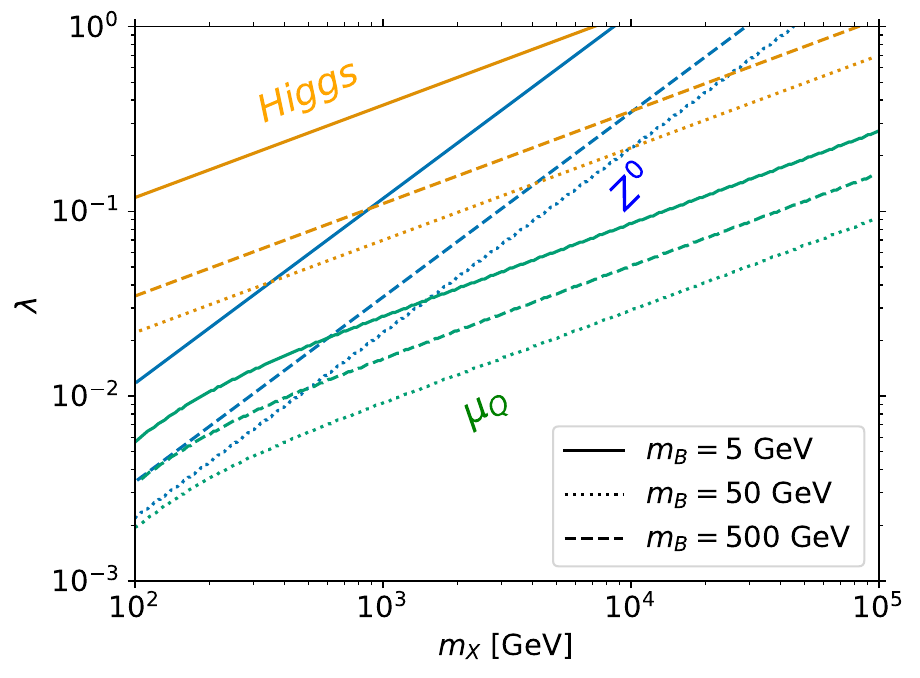}
 \caption{Direct detection upper limits on $\lambda$ (Eq.\ (\ref{xdecay})) versus $m_{X}$ for the Dirac $X$ mediator, assuming $N=3$ dark colors. The green, blue and yellow lines denote constraints from magnetic dipole moment ($\mu_Q$), $Z^0$ portal scattering ($\sigma_Z$) and Higgs portal scattering ($\sigma_H$), respectively. Different line styles distinguish values of the dark baryon mass $m_B$ as indicated.
 } 
 \label{fig:cs}
\end{center} 
\end{figure}

The contribution to direct detection signals from $Z$ exchange has been computed in Refs.~\cite{Goodman:1984dc,Essig:2007az} for DM particles with nonzero hypercharge. 
Accounting for the mixing and number of dark quark colors,
we find the cross section for scattering on nucleons
\be
    \sigma_Z \cong
    {(N\theta^2 G_F \mu_{nB})^2\over \pi}\,\left(1 - 1.08\,{Z\over A}\right)^2,
\ee    
where $\mu_{nB}$ is the nucleon-DM reduced mass, and $Z,A$ are the charge and atomic mass of the target nucleus.
For the numerical evaluation, we use $\sin^2\theta_W = 0.23$.

For the Higgs portal, whose amplitude is suppressed by only one power of $\theta$,  we estimate the $h$-$B$ coupling to be $N\lambda$.
The Higgs-nucleon coupling is $y_N \cong 0.3\, m_N/v$ \cite{Hoferichter:2017olk}, 
giving 
\be
 \sigma_H=\frac{4(N y_N\theta\lambda)^2\mu_{nB}^2}{\pi m_H^4} =\frac{y_N^2 N^2\lambda^4 \mu_{nB}^2}{2\pi \lambda_H^2 m_\chi ^2 v^2}\,.
 \label{Higgs_DiracX}
\ee
In the second form, $\theta$ and $m_H$ have been
eliminated in favor of the 
Higgs self-coupling $\lambda_H=0.13$ and its  vacuum expectation value (VEV) $v=246$~GeV. The
$Z^0$ and Higgs portals are constrained by direct detection results from Ref.\ \cite{PandaX-4T:2021x}, leading to the upper limits on $\lambda$ versus $m_{X}$ shown in Fig.~\ref{fig:cs}.
The $Z$ and $H$ exchange limits correspond to the blue and orange lines, respectively, for several choices of the DM particle mass
$m_B$.

The PandaX-4T experiment recently improved the limits
on magnetic dipole moment mediated scattering on 
protons~\cite{PandaX:2023}. 
The dark quark magnetic moment $\mu_Q$ was estimated in Eq.~\eqref{MDM}, and the contribution $\mu_Q \sim \theta^2 g_2^2 e/(16\pi^2 m_X)$ from the virtual $W$ diagram is subdominant except for $m_X\lesssim 400\,$GeV.
Taking the dark baryon moment to be $\mu_B = N \mu_Q$, the resulting constraints on $\lambda$ versus $m_X$ are shown in Fig.~\ref{fig:cs} as green lines, for the benchmark case of $N=3$.
We find that the magnetic moment constraints dominate the bounds on $\lambda$ over the whole range of DM and mediator masses considered.

\subsection{Scalar $X$ mediator}
\begin{figure}
\begin{center}
 \includegraphics[width=0.5\textwidth]{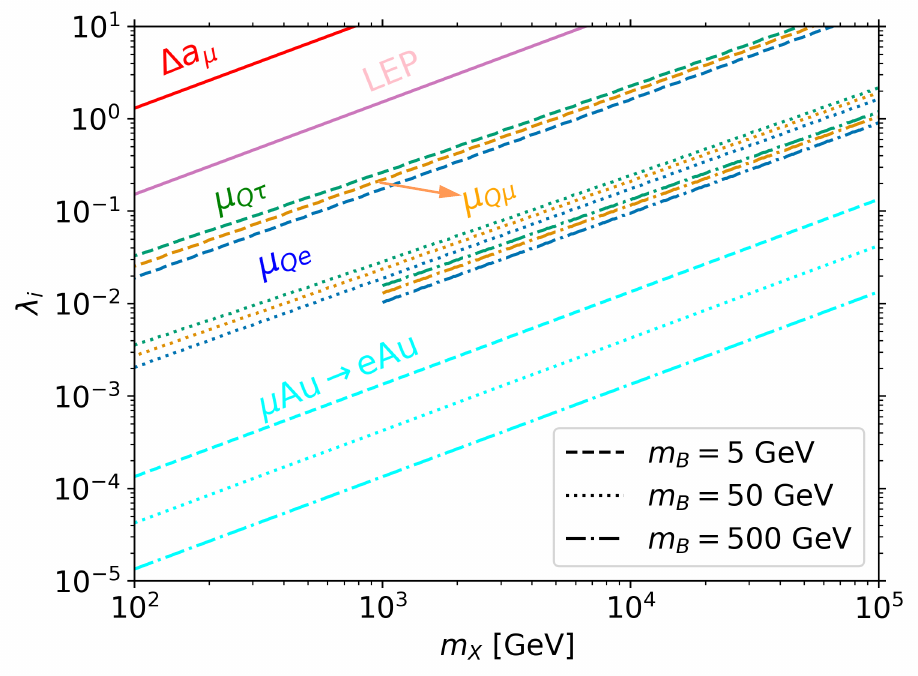}
 \caption{Constraints on the  $\lambda_{i}$ ($i=e,\mu,\tau$) portal couplings of Eq.~\eqref{xdecay} for the scalar $X$ mediator.   
 The red line shows the  constraint on $\lambda_\mu$ from $(g-2)_\mu$, see Eq.~\eqref{muonMDM}, and the purple line the one on $\lambda_e$ from LEP, see Eq.~\eqref{LEP}.    The lowest three curves (cyan) show limits on $\sqrt{\lambda_e\lambda_\mu}$
 from $\mu\to e\gamma$ in Au nuclei, as described by Eq.\ (\ref{mu2eg}),
 assuming $m_B = 3 m_Q$. The intermediate lines are upper limits from direct detection via the magnetic dipole moment $\mu_{Qe}$ (blue), $\mu_{Q\mu}$ (orange) and $\mu_{Q\tau}$ (green).
 $m_B$ is assumed to be less than $m_X/2$ to enforce a hierarchy of scales.
 }
 \label{fig:MDM-scalar X}
\end{center} 
\end{figure}

For the scalar $X$ variant, the  portal interaction (\ref{xdecay}) is to right-handed 
standard model leptons. The magnetic dipole moment of the dark quark generated from the $X$-lepton loop is given in Eq.\ (\ref{MDM}),
with the assumption $m_X \gg m_Q$.
In applying the experimental limits on dark matter MDM ~\cite{PandaX:2023}, we  assume that only one coupling $\lambda_i$ is turned on at a time, with $i=e,\mu,\tau$.  The resulting constraints are shown in Fig.\ \ref{fig:MDM-scalar X}. For comparison, the weaker limits on $\lambda_e$ from LEP 
(\ref{LEP}) and on $\lambda_\mu$ from the muon anomalous moment (\ref{muonMDM}) are also shown.

In the scalar $X$ model, there is no mass mixing between
$Q$ and a heavier state, hence no Higgs or $Z$ interactions are induced at tree level.  We neglect the $Z$ magnetic moment contribution since it is suppressed compared to the electromagnetic one.  An additional operator $\lambda'|X|^2|H|^2$ is allowed, which at one loop leads to a chirally-suppressed coupling of Higgs to $Q$ of order $N\lambda'\lambda_i^2 m_i v/(16\pi^2 m_X^2)$, where $m_i$ is the lepton mass.  The resulting constraints on $\lambda'\lambda_i^2$, even if saturating the perturbativity constraint on $\lambda'$, are quite weak relative to those from the dipole moment, even for values of $\lambda'$ that
saturate perturbative unitarity.

\subsection{Kinetically mixed composite vector
exchange}
We have identified three possible composite vector states that kinetically mix with the photon, and couple vectorially to the dark baryon.
The mixing and coupling parameters $\epsilon_v$ and $g_v$ are summarized in Table \ref{tab2} for the three states $v = \tilde A^\mu$, $\omega_\mu$ and ${\cal G}^\mu$.  The cross section for the vector-mediated DM scattering on protons is 
\be
    \sigma_{v}=\frac{(\epsilon_{v} g_{v} e)^2
    \mu_{nB}^2}{\pi {m_{v}}^4}\,,
\ee
where $\mu_{nB}=m_nm_B/(m_n+m_B)$ is the nucleon-$B$ reduced mass.  A similar expression 
holds for DM-electron scattering, by replacing
$\mu_{nB}$ with the electron-$B$ reduced mass.
In the latter case we assume that that $m_v \gg \alpha m_e$, which is the momentum scale of typical electron interactions of gaseous atomic  and semiconductor detectors.

\begin{figure}
\begin{center}
 \includegraphics[width=0.5\textwidth]{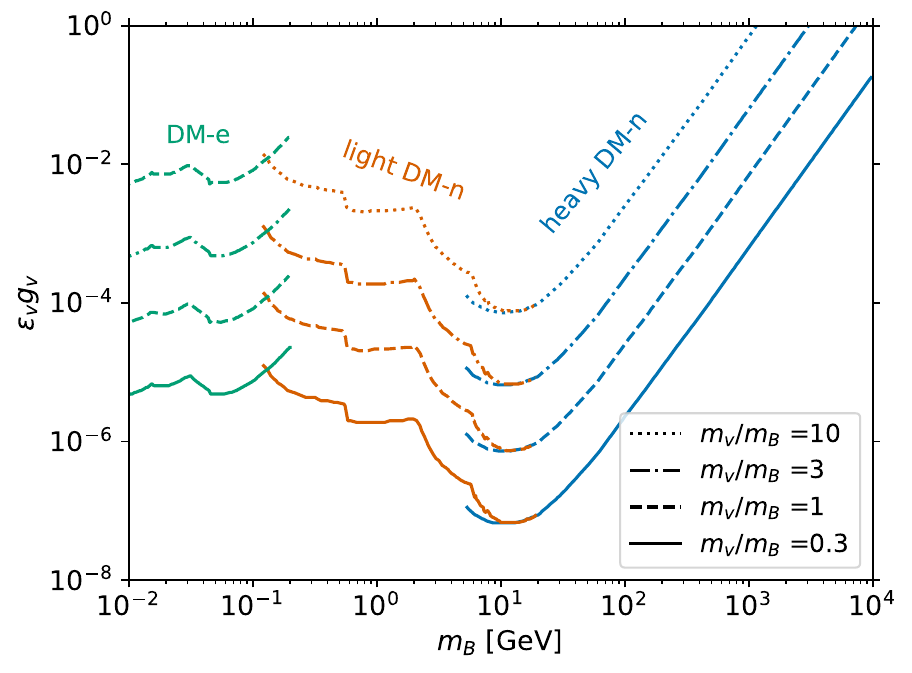}
 \caption{Direct detection constraints on $\epsilon_{v} g_{v}$ (kinetic mixing times coupling to dark baryon) versus $m_{B}$, from exchange of a   kinetically mixed composite vector boson. Different linestyles distinguish choices of the ratio of dark vector to dark baryon mass $m_v/m_B$, as labeled. Green, yellow, and blue colors represent constraints from heavy DM-nucleon~\cite{PandaX-4T:2021x}, light DM-nucleon~\cite{Xenon2018,Xenon-light,Xenon-Migdal}, and DM-electron~\cite{Xenon-light,Panda-e-2023,sensei-e,DarkSide50}\cite{DAMIC-M_e} cross section, respectively.}
 \label{fig:Acontour}
\end{center} 
\end{figure}

\begin{figure*}[t]
\begin{center}
\centerline{{\includegraphics[width=0.3\textwidth]{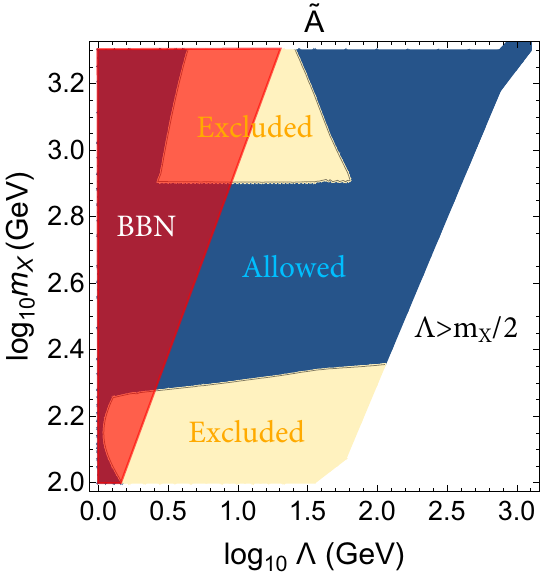}}\hfil
\includegraphics[width=0.3\textwidth]{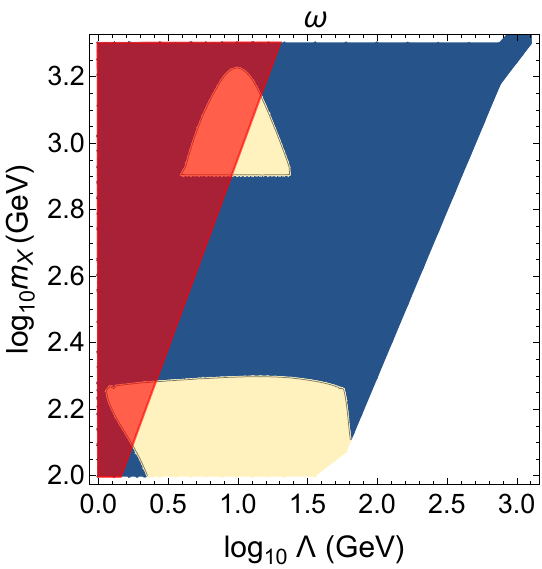}
\hfil
\includegraphics[width=0.3\textwidth]{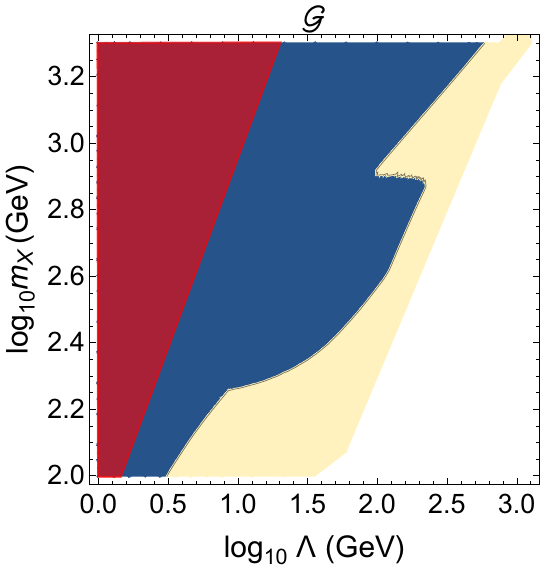}}
 \caption{Allowed (blue) and excluded (yellow) regions of $m_X$ versus $\Lambda$ from direct detection constraints on exchange of kinetically mixed vector particles $\tilde A^\mu$ (left), $\omega^\mu$
 (center) and ${\cal G}^\mu$ (right).  
 Red regions are excluded by the BBN constraint
 (\ref{gbdecconst}) from dark glueball decays, and white regions are excluded by the need of $\Lambda<m_X/2$.
 See text for details.
 }
 \label{fig:vector_comp}
\end{center} 
\end{figure*}

Combining the results from experimental searches for proton- and electron-DM scattering, we 
obtain constraints on the product 
$\epsilon_{v} g_{v}$ versus DM mass $m_B$, as
shown in Fig.~\ref{fig:Acontour}.  The experimental limits for scattering on protons
come from PandaX-4T~\cite{PandaX-4T:2021x} and XENON 1T~\cite{Xenon2018,Xenon-light,Xenon-Migdal}, and for scattering on electrons from XENON1T-S2~\cite{Xenon-light}, PandaX-4T~\cite{Panda-e-2023}, SENSEI~\cite{sensei-e}, DarkSide-50~\cite{DarkSide50} and DAMIC-M~\cite{DAMIC-M_e}. 

The most generic particle mediating this interaction is the $1^{--}$ vector glueball, since its low-energy interactions are independent 
of any dimensionless couplings present in the UV model; recall that the running dark gauge coupling is determined by $\Lambda/m_X$.  
On the other hand, it is the most highly suppressed contributor in the small parameter
$\Lambda/m_X$, with $\epsilon_{\cal G}$ scaling as $(\Lambda/m_X)^4$.  Hence, it can be subdominant to $\tilde A^\mu$ exchange if the combination $y_x y_q$ (or $ y_q\mu_x/m_X$) is
sufficiently large. 

To illustrate the complementarity of the various vector exchanges, we can exclude regions of the $m_X$-$\Lambda$ plane using direct detection, by making assumptions about the values of other relevant masses and couplings.  The constraints in Fig.\ \ref{fig:Acontour} require specifying the ratio $m_v/m_B$.  
Taking $m_\Phi\sim m_Q$ gives $m_{\tilde A}/m_B = 1/N$, $m_\omega/m_B = 2/N$ and $m_{\cal G}/m_B = 9/[N(1 + m_Q/\Lambda)]$.  We take $N=3$ for definiteness.  To satisfy LHC constraints, we assume $m_Q(\Lambda)$ saturates the bound (green curve) of Fig.\ \ref{fig:LHC_X0} (left) in the case of Dirac $X$.  To fix the strength of $\omega$ exchange, we take $\lambda(m_X)$ to follow the solid green curve shown in Fig.\ \ref{fig:cs}.
Similarly, to fix $\epsilon_{\tilde A}g_{\tilde A}$
we assume $y_x = y_q = 1$.  The resulting excluded regions are shown in yellow in Fig.\ \ref{fig:vector_comp}, for the Dirac $X$ case. The $\tilde A^\mu$ and $\omega^\mu$ give similar constraints, while the glueball ${\cal G}^\mu$ is
distinctive.  In all cases, the weakened limits at intermediate $m_X$ values reflect the fact that LHC constraints are strongest in this region, and push the dark baryon mass to higher values where the direct detection constraints become weaker.

\section{Lepton flavor constraints}
\label{sec:lfuv}

In the case of a scalar $X$ mediator,
the $\lambda_i \bar Q X e_{R,i}$ interaction in Eq.\ (\ref{xdecay}) gives rise to processes that can violate lepton flavor universality or conservation.
One such effect is a negative contribution to the muon anomalous magnetic moment \cite{Queiroz:2014zfa},\be
\Delta a_\mu = -{\lambda_\mu^2\, m_\mu^2\over 96\pi^2\, m_X^2 }\,,
\ee
which exacerbates the 
tension ($\Delta a_\mu = 251\times 10^{-11}$
\cite{Workman:2022ynf}) between experiment and the SM prediction.
Conservatively assuming that this tension is due to hadronic uncertainties, we estimate that $\Delta a_\mu > -200\times 10^{-11}$ at 3$\sigma$, leading to the bound
\be
    {m_X\over \lambda_\mu} > 77\,{\rm GeV}\,.
\label{muonMDM}
\ee
Analogous bounds from the electron anomalous moment are much weaker, because of the chiral suppression.

We also consider constraints on flavor-changing neutral current (FCNC) processes induced by $X$. The current best limit on lepton flavor-violating muon decays is set by the Mu to E Gamma (MEG) experiment, which finds BR($\mu\rightarrow e\gamma$)$< 4.2\cdot10^{-13}$ \cite{meg2016}. The transition magnetic dipole moment is given in our model by
\be
\mu_{e\mu} \sim \frac{e\lambda_e\lambda_{\mu}m_Q}{16\pi^2m_X^2}\,,
\ee
leading to a constraint on the product of couplings
$\lambda_{\mu}\lambda_{e} \lesssim 10^{-4}\left(m_X/{\rm TeV}\right)^2({\rm GeV}/{m_Q}).$

However, one can obtain stronger bounds from muon-to-electron conversion in muonic gold, with BR($\mu\, {\rm Au}\rightarrow e\,{\rm Au}$)$<7 \cdot 10^{-13}$ \cite{SINDRUMII:2006dvw}. The dipole operators in the general lepton flavor violating (LFV) lagrangian give 
\cite{DAVIDSON2019380}
\be
{\rm BR}(\mu\, {\rm Au}\rightarrow e\, {\rm Au})=\frac{2G_F^2m_{\mu}^5C_{D,R}^2D^2}{\Gamma_{\rm capt}}\,,
\ee
where $C_{D,R}$ is the Wilson coefficient for the right-handed dipole operator
$\bar e_L\sigma^{\mu\nu}\mu_R F_{\mu\nu}$
and $D$ is a nuclear overlap integral. With the measured muon capture rate \cite{Suzuki:1987jf} and $D=0.189$ \cite{Kitano:2002mt} for gold, we derive $C_{D,R}<10^{-9}$, which gives $\mu_{e\mu}<10^{-15}$ GeV$^{-1}$, which is two orders of magnitude more stringent than the previously discussed one. This results in a bound
\be
\lambda_e\lambda_{\mu} \lesssim 10^{-6}\left({m_X\over {\rm TeV}}\right)^2 \frac{{\rm GeV}}{m_Q}\,.
\label{mu2eg}
\ee
If one assumes lepton flavor universality, so that all $\lambda_i$ are equal, this constraint is more stringent than those coming from direct detection in Fig.~\ref{fig:MDM-scalar X}, taking $m_Q = m_B/3$. 
If $m_Q\ll\Lambda$ so that the dark baryon gets its mass mostly from confinement, the constraint
(\ref{mu2eg}) can however become weaker than the direct detection ones.

\begin{figure*}[t]
\begin{center}
\centerline{
 \includegraphics[width=0.58\textwidth]{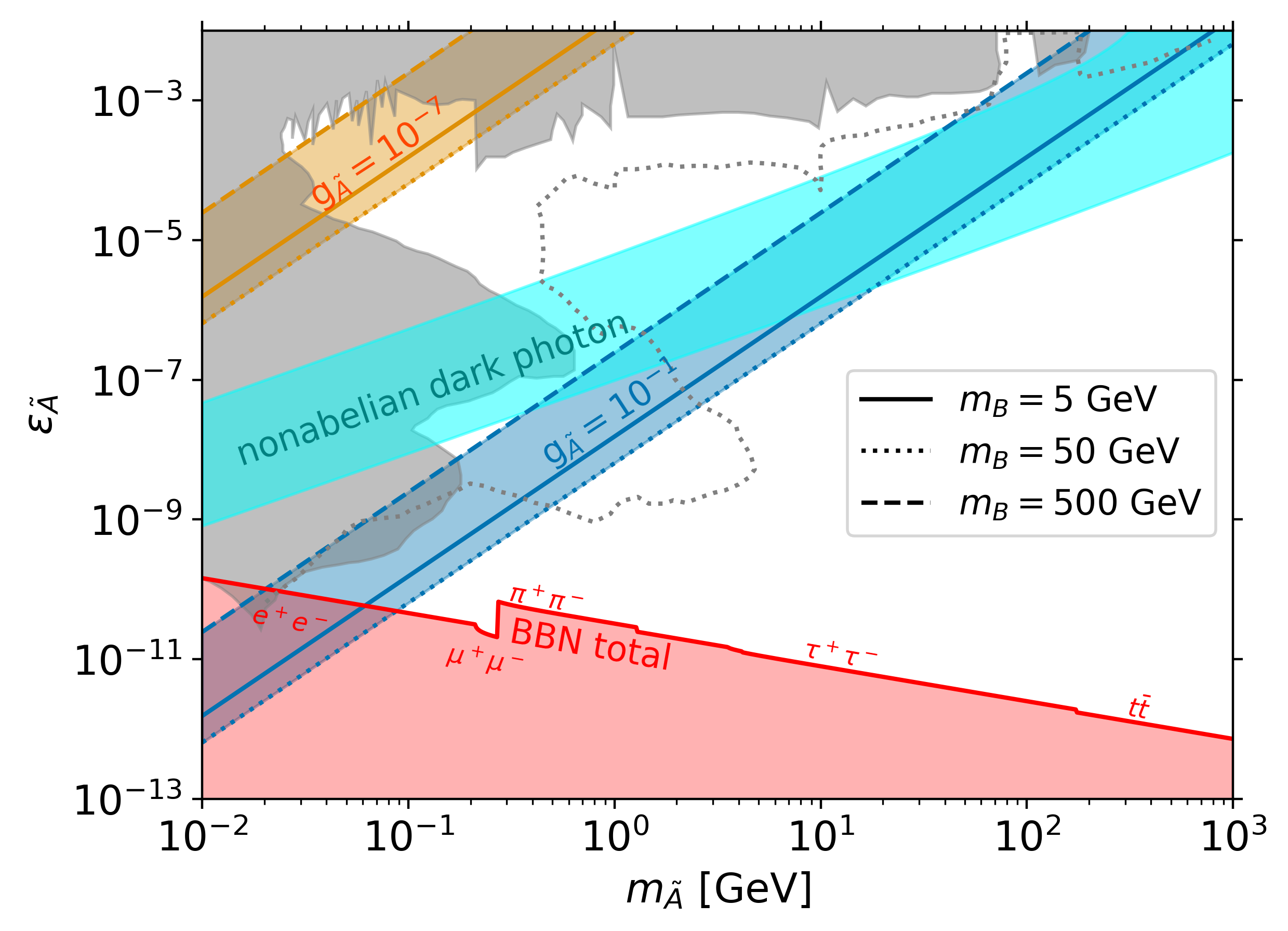}
 \includegraphics[width=0.42\textwidth]{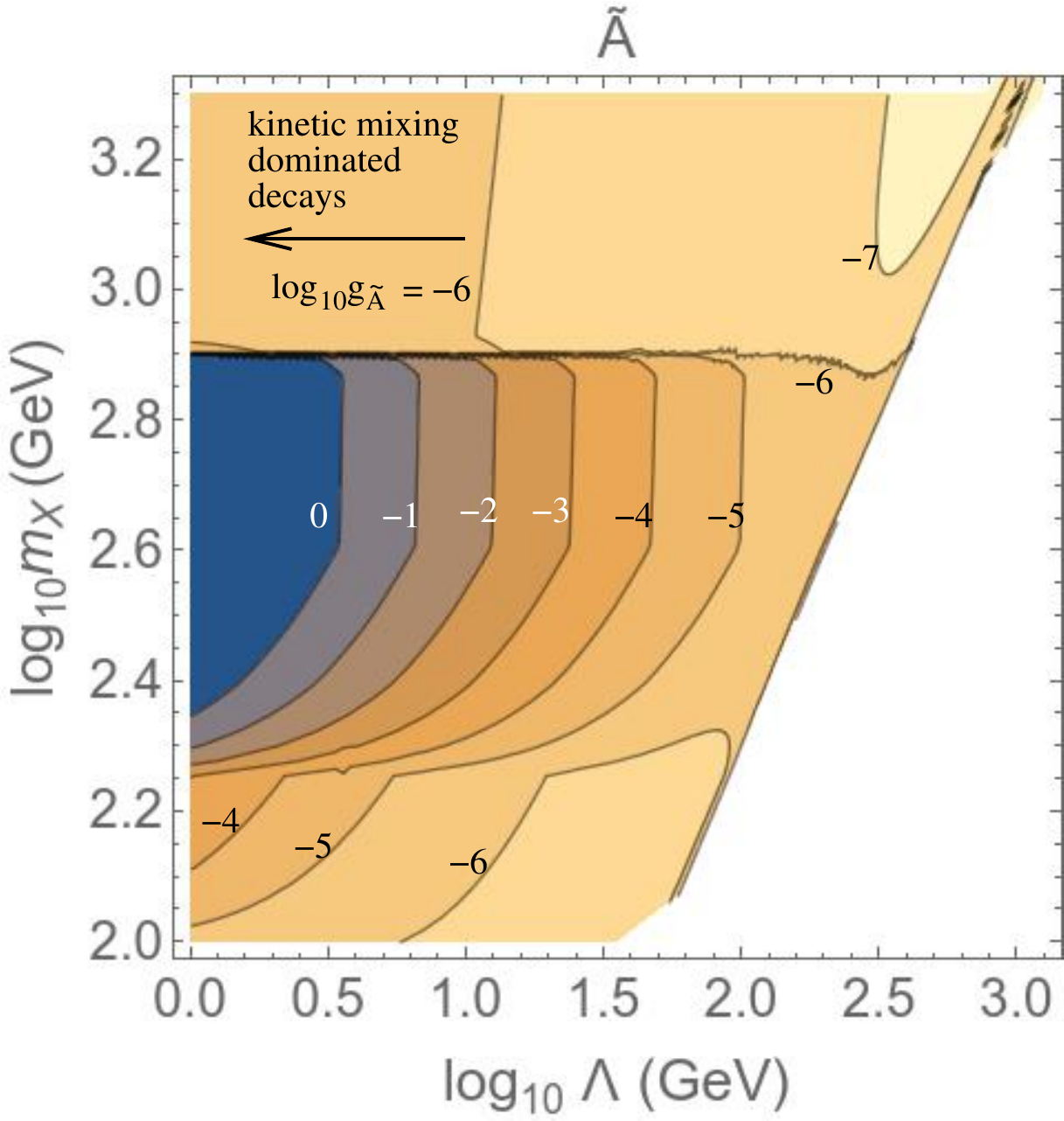}
 }
 \caption{\emph{Left:} BBN lower bound on kinetic mixing of $\tilde A$ (solid red curve, with dominating final states labeled) versus its mass. Laboratory~\cite{Fabbrichesi:2020wbt} (and references therein) and supernovae~\cite{SN_dark_photon} excluded regions are grey shaded, while theoretical forecasts \cite{Agrawal:2021dbo} are shown as the grey dotted curve. Orange or blue shaded regions are respective upper bounds from direct detection~\cite{PandaX-4T:2021x} assuming $g_{\tilde{A}} = 10^{-7}$ or $10^{-1}$, with $m_B$ ranging from 5 to 500 GeV. The cyan shaded region shows prediction of our model on where the dark photon lives (see text for details). \emph{Right:} regions of $m_X$ versus $\Lambda$  in which kinetic mixing dominates over mass mixing for $\tilde A$ decays are to the left of the curves, labeled by $\log_{10} g_{\tilde A}$.} 
 \label{fig:Atdecay}
\end{center} 
\end{figure*}

\section{Cosmological Implications}
\label{sec:cosmo}

In our scenario, the $X$ mediator is expected to be in equilibrium with the SM particles at temperatures
$T \gg m_X$, by virtue of its weak hypercharge.  
The dark sector particles will then also equilibrate, since the gauge coupling $g'(m_X)$ cannot be too small.  Therefore the relative abundance of dark sector particles is only suppressed by the entropy dilution produced during the QCD phase transition, and this requires any
long-lived particles to decay with lifetimes
$\lesssim 0.2\,$s to avoid disrupting Big Bang Nucleosynthesis (BBN) \cite{Kawasaki_2018}.  
For a dark photon that decays via kinetic mixing, like our composite $\tilde A^\mu$ state, this leads to a \emph{lower} limit on the kinetic mixing parameter.
This is unlike the case for typical dark photons, for which the relic density is assumed to arise from the kinetic mixing itself through freeze-in~\cite{Fradette:2014sza,Berger:2016vxi,Li:2020roy}, leading to very different limits.

The least model-dependent constraint comes from the decays of dark glueballs, leading to a minimum value of $\Lambda$ for given $m_X$~\cite{Forestell:2017wov}.  Demanding that the lowest glueball lifetime from Eq.~\eqref{gbdecay} be less than $0.2\,$s we find 
\be
    m_X < 69\,{\rm  GeV}\left(\Lambda\over{\rm GeV}\right)^{1.13}\left(N\over 3\right)^{1/4},
    \label{gbdecconst}
\ee 
from a numerical fit for $N=3$.  (The exponent in Eq.~\eqref{gbdecconst} is less than $9/8$ due to the running of $\alpha$'.) 

The next least model-dependent process is
the decay of $\eta$ to two photons, 
induced by the operators in Eqs.~\eqref{meson-decays} and \eqref{divj5}.
In analogy to the rate for $\pi^0\to \gamma\gamma$ in the standard model, we find
\be
    \Gamma(\eta\to\gamma\gamma) = {N^2\alpha^2\alpha'^4 \ln^2(m_X/m_\eta) m_\eta^3\over 64\pi^7 f_\eta^2}
    \label{etadecay},
\ee
where we have taken $f_\eta\sim\Lambda$.
This rate depends on the additional parameter $m_Q$ through $m_\eta$, which we fix to its minimum value as a function of $m_X$ from the LHC constraints of Section \ref{sec:coll}.  Since this decay is much faster than that of the glueball, it adds no constraint beyond Eq.~\eqref{gbdecconst} to the parameter space shown in Fig.~\ref{fig:vector_comp}.  Excited glueball states typically have faster radiative decays into lower ones, making~\eqref{gbdecconst} the the most relevant constraint. The decay  $\omega^\mu\to\eta\gamma$ is also fast and imposes no further restrictions on the model.

The $\tilde A^\mu$ composite vector can decay 
to three photons via its mixing with $\omega^\mu$, through 
$\tilde A^\mu \to \eta^*\gamma\to 3\gamma$,
where $\eta^*$ is virtual if $m_{\tilde A} < m_\eta$.  
The rate for this process is approximately
given by Eq.~\eqref{etadecay} multiplied by
$\theta_{\tilde A}^2$ (see Eq.~\ref{thetaA}),
replacing $m_\eta\to m_{\tilde A}$, and including the additional factor $\alpha m_{\tilde A}^4/(4\pi m_\eta^4)$ for the emission of the first photon and the virtual
$\eta$ propagator.  In addition, $\tilde A$ can decay to SM charged pairs $f\bar f$ through kinetic mixing, with the rate \cite{Fabbrichesi:2020wbt}\footnote{For decays into hadrons, one uses the rate for muons times the ratio $R=\sigma_{e^+e^-\to {\rm had}}/
\sigma_{e^+e^-\to \mu^+\mu^-}$.}
\be
    \Gamma(\tilde A \to f\bar f)=\frac{(Q_f\epsilon_{\tilde A}e)^2\left(m_{\tilde A}^2-4m_q^2\right)^{1/2}\left(1+2\frac{m_q^2}{m^2_{\tilde A}}\right)}{12\pi m_{\tilde A}^2}\,.
\ee
If the hadronic 3-photon process is too slow, then the kinetic-mixing-induced decays must be fast enough to satisfy BBN constraints.  
Fig.~\ref{fig:Atdecay} (left) shows the lower bounds on $\epsilon_{\tilde A}$ such that the lifetime constraints from Ref.\ \cite{Kawasaki_2018} are satisfied, depending on the accessible final states.  Since BBN is more sensitive to the injection of hadrons, there is a noticeable jump in the constraint at the threshold for decay into $\pi^+\pi^-$.

Assuming $m_\Phi\sim\Lambda$ for definiteness, to determine
$m_{\tilde A}\sim 2\Lambda$, 
Fig.~\ref{fig:Atdecay} (right) shows the regions of $m_X$ versus $\Lambda$ where the
$\tilde A\to 3\gamma$ channel is subdominant to
$\tilde A\to f\bar f$: they lie to the left
of the curves, which are labeled by the assumed value of $g_{\tilde A}$, the coupling of $\tilde A$ to the dark baryon, see Eq.~\eqref{epsilon_eff}. 
This coupling is relevant for the direct detection limits, shown as diagonal lines in the $m_{\tilde A}$-$\epsilon_{\tilde A}$ plane (left panel of Fig.~\ref{fig:Atdecay}), and it also appears in the $\theta_{\tilde A}$ mixing angle.  For $g_{\tilde A}\sim 10^{-7}$, for example, kinetic mixing dominates over most of the parameter space, so that the solid red lower bound on $\epsilon_{\tilde A}$ in the left plot applies, and it is consistent with the direct detection bound.   

On the other hand, for $g_{\tilde A}\sim 1$, the kinetic mixing lower bound is only relevant for $\Lambda \lesssim 3\,$GeV and intermediate values of $m_X \in (220,800)\,$GeV that are restricted by LHC searches.  For $g_{\tilde A}\sim 1$, the left plot of
Fig.\ \ref{fig:Atdecay} shows that BBN plus direct detection excludes $m_{\tilde A}\lesssim 1$\,GeV.  

An interesting distinction from generic dark photon models is the prediction that the kinetic mixing of the various vectors can be correlated with their mass, from Tables \ref{tab1} and \ref{tab2}.  For the $\tilde A$ state, we make the  simplifying assumption that $m_\Phi\sim\Lambda$ so that
$m_{\tilde A}\sim 2\Lambda$.  The predicted region is shown in Fig.\ \ref{fig:Atdecay} (left) for representative choices $y_x=1$ and valid for 
$m_X$ ranging from 0.2 to 10 TeV, as cyan shaded band.

The complementarity between dark photon and direct detection searches is an interesting feature of our scenario.
A significant area of the region highlighted in cyan in Fig.~\ref{fig:Atdecay} will be covered by upcoming and proposed terrestrial searches for dark photons~\cite{Beacham:2019nyx,Fabbrichesi:2020wbt,Agrawal:2021dbo}, while at the same time being accessible to DM direct detection searches.
Thus, a simultaneous and compatible signal found in this region would be a strong indication of a dark photon of composite origin in a confining dark sector. 

\section{Conclusions}
\label{sec:conc}
Massive dark photons are a popular subject of study with many applications in beyond the standard model scenarios.
In this work, we have highlighted that they need not get their mass from the Higgs or St\"uckelberg mechanisms, as is usually assumed, but may instead arise from compositeness in a confining SU(N)$'$ dark gauge sector. 
Indeed, such models typically contain several vector states that can kinetically mix with the SM photon: mesons, glueballs, and other composites depending upon the field content of the dark sector.

It is generic that if a heavy mediator particle $X$ exists that carries both SU(N)$'$ quantum numbers and standard model hypercharge, all of these dark photons will acquire kinetic mixing.
Assuming the dark matter to be some dark hadronic state, this provides a means for dark matter-proton scattering for direct detection. 
A further generic requirement is that $X$ should have additional interactions involving SM states allowing it to decay, as otherwise exotic charged relics would have been observed in the universe. 
This typically introduces additional possibilities for direction detection beyond the kinetic mixing portal,
leading to complementary constraints on the model parameters.

The additional portals can also make the $X$ mediator discoverable in particle colliders. 
LHC searches exclude $m_X \lesssim 800\,$GeV unless the dark quark $Q$, which contributes to missing energy in the $X$ decay, is sufficiently heavy (typically several hundred GeV). 
For heavier $X$ particles, $m_Q$ could have a negligible mass since it will hadronize into states of mass $\sim\Lambda$. 
On the other hand, direct detection can be sensitive to $m_Q$ by itself since the loop-induced magnetic dipole moment of $Q$ is chirally suppressed if $X$ is a scalar.  This is an example of the complementarity of different experimental constraints.

The extra portals can also lead to new low-energy effects such as lepton flavor violation. For a scalar mediator, muon to electron conversion in nuclei provides the most sensitive current probe if $X$ couples to both flavors. 
For couplings of $X$ to a single lepton flavor, we find direct dark matter searches (via the dipole moment interaction) to give the strongest constraints on the model parameters.

Since the mediator equilibrates with SM particles in the early universe, the abundance of dark sector particles is generically comparable to that of the SM ones.
Thus, the new dark states must decay sufficiently early so as not to affect the predictions of big bang nucleosynthesis. 
In the minimal models, where only dark vector mesons or glueballs are kinetically mixed, the latter exclude regions of small $\Lambda$, depending on $m_X$, while the former generically decay fast, since the collider constraints force them to be relatively heavy.
Interestingly, BBN gives rise to a lower bound on the kinetic mixing of the exotic $\tilde A^\mu$ vector.
This is a novel feature that is not present in generic dark photon models, and is due to the distinctive mechanism of populating the states in the confining model in the early universe.

The resulting phenomenology is quite rich, with many connections between the various experimental constraints.
As an example, Fig.~\ref{fig:Atdecay} showcases that a simultaneous direct detection signal and terrestrial discovery of the nonabelian dark photon is possible in upcoming and proposed experimental setups.
This optimistic situation is what one would hope for in terms of being able to pin down the detailed nature of particle dark matter and its possibly accompanying hidden sector.

Of necessity, we have estimated hadronic matrix elements in terms of the confinement scale $\Lambda$ using dimensional analysis, sometimes supplemented by large-$N$ or quark model arguments, or comparison to QCD.  These estimates are needed to connect fundamental parameters of the model to the low-energy observables. In future studies, computing these nonperturbative quantities on the lattice~\cite{DeGrand:2019vbx} would enable more accurate predictions.

\bigskip {\bf Acknowledgments.} We thank Tom DeGrand, Guy Moore, Joerg Jaeckel, and Ethan Neil for helpful discussions. This research
was supported by the Natural Sciences and Engineering Research Council (NSERC) of Canada.

\bibliography{ref}
\bibliographystyle{utphys}

\end{document}